\documentclass[11pt,preprint]{aastex}

\renewcommand{\deg}{\mbox{$^\circ$}}

\newcommand{\be}{\begin{equation}}
\newcommand{\ee}{\end{equation}}

\newcommand{\bea}{\begin{eqnarray}}
\newcommand{\eea}{\end{eqnarray}}

\newcommand{\um}{\hbox{$\mu$m}}

\newcommand{\nWmmsr}{\hbox{nW m$^{-2}$ sr$^{-1}$}}

\newcommand{\kJysr}{\hbox{kJy sr$^{-1}$}}

\newcommand{\etal}{{\it et al.\/}}
\newcommand{\vs}{{\it vs.\/}}

\newcommand{\Kband}{\mbox{$K$-band}}

\newcommand{\Jband}{\mbox{$J$-band}}

\newcommand{\KkJy}{\hbox{$14.8 \pm 4.6$~\kJysr}}
\newcommand{\KnW}{\hbox{$20.2 \pm 6.3$~\nWmmsr}}
\newcommand{\JkJy}{\hbox{$12.0 \pm 6.8$~\kJysr}}
\newcommand{\JnW}{\hbox{$28.9 \pm 16.3$~\nWmmsr}}

\begin{document}

\title{DIRBE Minus 2MASS: Confirming the Cosmic Infrared Background at
2.2 Microns}

\author{Edward L. Wright} \affil{Department of Physics and
Astronomy, University of California, Los Angeles, CA  90095-1562}  
\email{wright@astro.ucla.edu}

\begin{abstract}
Stellar fluxes from the 2MASS catalog are used to remove the contribution
due to Galactic stars from the intensity measured by DIRBE in four
regions in the North and South Galactic polar caps.
After subtracting the interplanetary and Galactic foregrounds, 
a consistent residual intensity of \KkJy\ or \KnW\ at 2.2 $\mu$m is found.
At 1.25~$\mu$m the residuals show more scatter and are a much smaller
fraction of the foreground, leading to a weak limit on the CIRB of
\JkJy\ or \JnW\ (1 $\sigma$).
\end{abstract}


\section{Introduction}

The Diffuse InfraRed Background Experiment (DIRBE) on the COsmic
Background Explorer ({\sl COBE}, see \citet{Bo92}) observed the
entire sky in 10 infrared wavelengths from 1.25 to 240 \um.  
\citet{HAKDO98} discuss the determination of the Cosmic InfraRed
Background (CIRB) by removing foreground emission from the DIRBE data.  
This paper detected the CIRB at 140 and 240 \um, but only gives upper
limits at shorter wavelengths.  From 5 to 100 \um, the zodiacal light
foreground due to thermal emission from interplanetary dust grains is so
large that no reliable estimates of the CIRB can be made from a position 1
AU from the Sun \citep{KWFRA98}.  In the shorter wavelengths from 1.25 to
3.5 \um, the zodiacal light is fainter, but uncertainties in modeling the
foreground due to Galactic stars are too large to allow a determination of 
the CIRB \citep{AOWSH98}.  Recently, \citet{GWC00} removed the Galactic star
foreground by directly measuring the stars in a $2\deg\times2\deg$ box
using ground-based telescopes and then subtracting the stellar
contribution from the DIRBE intensity on a pixel-by-pixel basis.  This
field, a DIRBE dark spot, was selected using DIRBE data to have a minimal
number of bright Galactic stars.  In addition, \citet{WR00} used a
histogram fitting method to remove the stellar foreground from the DIRBE
data in a less model-dependent way than that used by \citet{AOWSH98}.
\citet{GWC00} and \citet{WR00} obtained consistent
estimates of the CIRB at 2.2 and 3.5 \um.  With the recent 2$^{nd}$
incremental release of 2MASS data, it is now possible to apply the direct
subtraction method of \citet{GWC00} to four additional DIRBE dark spots 
scattered around the North and South Galactic polar caps.

\citet{KO00} have claimed a detection of the fluctuations of the CIRB.
\citet{KO00} also give the range 0.05 to 0.1 for the ratio of the
fluctuations in the DIRBE beam to the mean intensity for the CIRB.
But this ratio and fluctuation combine to give a range of CIRB values
that is incompatible with the \citet{HAKDO98} upper limits on the
CIRB, especially at 1.25 \um.
Furthermore, the claimed cosmic fluctuations are larger than the residuals
in the DIRBE$-$2MASS fits presented in \S\ref{sec:analysis}.
In this paper, \citet{KO00} is treated as an upper limit on the CIRB which
is compatible with previous limits and the results found here.
\citet{Wr01} will discuss the possible cosmic fluctuation signal in
the DIRBE$-$2MASS residuals in more detail.

\section{Data Sets}

The two main datasets used in this paper are the DIRBE maps and the 2MASS
point source catalog (PSC).

The DIRBE weekly maps were used: DIRBE\_WKnn\_P3B.FITS for
$04 \leq \mbox{nn} \leq 44$.   These data and the very strong no-zodi
principle described by \citet{Wr97} were used to derive a
model for the interplanetary dust foreground that is described
in \citet{Wr98} and \citet{GWC00}.  The zodiacal light model was then
subtracted from each weekly map, and the remainders were averaged
into mission averaged, zodiacal subtracted maps.  At 1.25 and 2.2 \um,
no correction for interstellar dust emission is needed \citep{AOWSH98}.
The pixels in these mission averaged, zodiacal subtracted maps provide the
DIRBE data, $D_i$.

DIRBE has a $0.7^\circ\times 0.7^\circ$ square beam with a diagonal 
of $1^\circ$.  Thus a thick buffer ring is needed around any studied field
to keep bright stars outside the field from influencing
the measured DIRBE intensity. 
One can minimize the resulting inefficiency by studying fields with a large
area:perimeter ratio.  Large circular regions have the largest area:perimeter
ratio, and for circles as large as $3^\circ$ diameter it is still possible to
find fields that have no stars brighter than the 2MASS saturation limit.
To find such fields,
a list of DIRBE dark spots was generated by smoothing the 
zodi-subtracted 3.5 \um\ map with a kernel given by 
\be
\log_2 W = \frac{-|\hat{n}-\hat{n}^\prime|^2}
		{3(0.03023^2-|\hat{n}-\hat{n}^\prime|^2)}
\ee
where $\hat{n}$ and $\hat{n}^\prime$ are unit vectors.  This kernel and 
all of its derivatives are continuous, and it vanishes identically for
radii greater than $2 \sin^{-1}(0.03023/2) = 1.732^\circ$.
The FWHM is 3\deg.  The 20 faintest spots of the smoothed map 
in the Northern Galactic Hemisphere and the 20 faintest spots in the Southern 
Galactic Hemisphere were then located.  The darkest spot is in the Northern
Hemisphere and was studied by \citet{GWC00} using ground-based data.

\begin{figure}[tbp]
\plotone{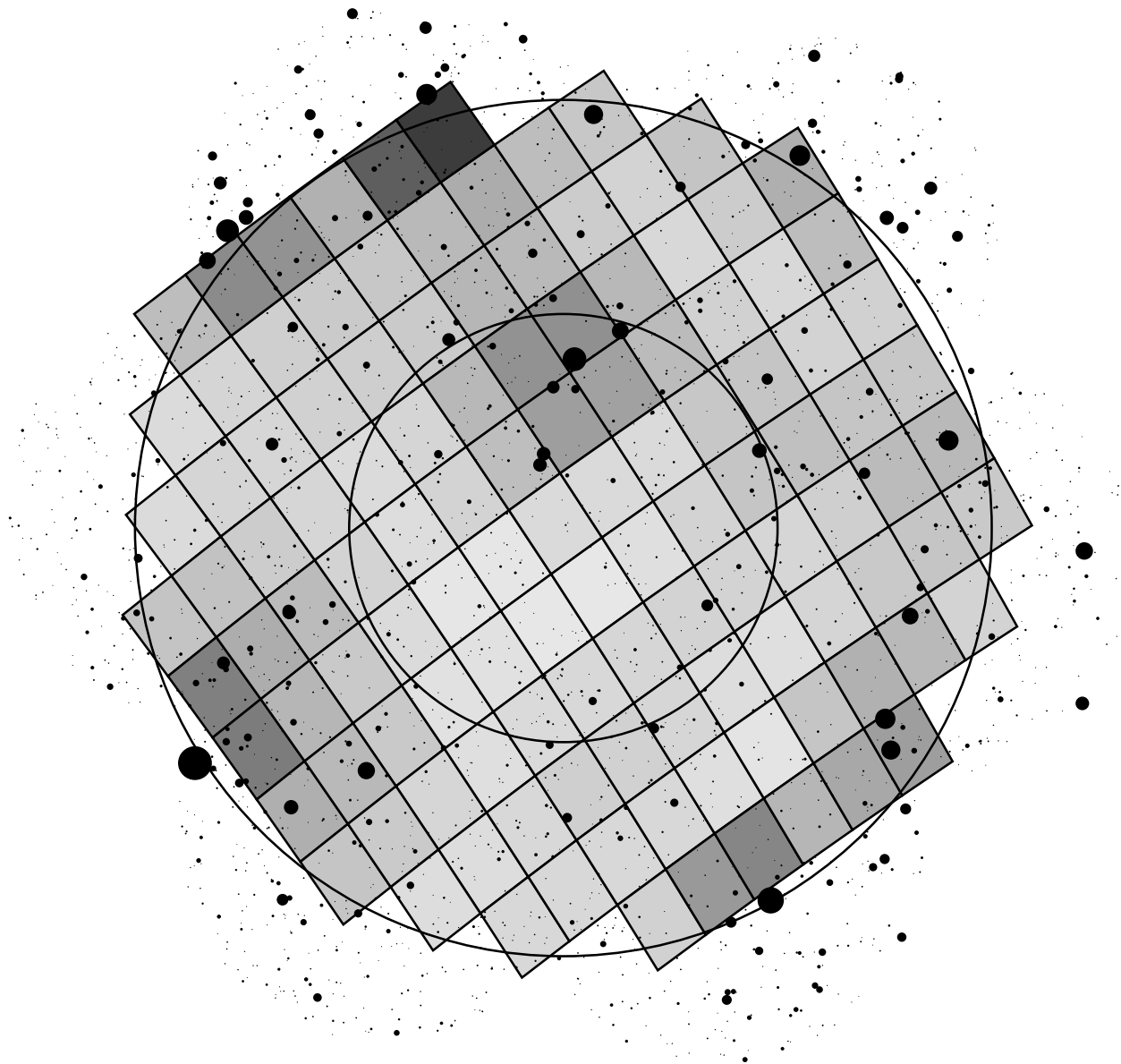}
\caption{The
field near the dark spot at $(l,b)$ = $(107.7\deg, 57.7\deg)$.
The grid shows the DIRBE pixel outlines, while the grayscale
indicates the DIRBE flux at 2.2 \um.  The filled black circles
are the 2MASS stars with $K_s < 12.5$, with the circle area
proportional to the flux.  The large open circles show a 1 and
2\deg\ radius around the dark spot.
\label{fig:CS13-map}}
\end{figure}

The 2MASS data were obtained over the WWW between 16 Mar 2000
and 23 Mar 2000.  The IRSA interface was used to search the catalog.
This allows a maximum search area of a 1\deg\ radius circle, which is too
small for a comparison to the DIRBE data taken with a 0.7\deg\ beam.
Therefore, a total of seven cone searches in a pattern consisting of a 
hexagon plus the central position were made around each DIRBE dark spot.  
These searches were restricted to stars brighter than $K_s = 14$.
The seven resulting files were combined by stripping the table headers, 
concatenating, sorting, and then using the UNIX uniq filter.  
If there are no gaps in the 2MASS coverage
near a dark spot, this yields a catalog that covers a 2\deg\ radius circle
around the dark spot plus six small ``ears.''  Only stars within the circle
were used in this analysis.
However, there are usually gaps in the 2MASS coverage.
Searching the 20 darkest spots in each of the Galactic polar caps produced
only four usable fields out of 40 candidates.  
These are listed in Table \ref{tab:K}.
$\beta$ is the ecliptic latitude.  The combined catalogs for each field were
converted into star charts which were checked for missing 2MASS data.
Two of these four fields have coverage gaps near the edge of the 2\deg\ radius
circle, and thus have $r < 2\deg$ and a smaller but still useful number of 
pixels.  Figure \ref{fig:CS13-map} shows the 2MASS catalog stars 
superimposed on the DIRBE map for the dark spot at 
$(l,b) = (107.7\deg, 57.7\deg)$.

\section{Analysis}
\label{sec:analysis}

\begin{figure}[tbp]
\plotone{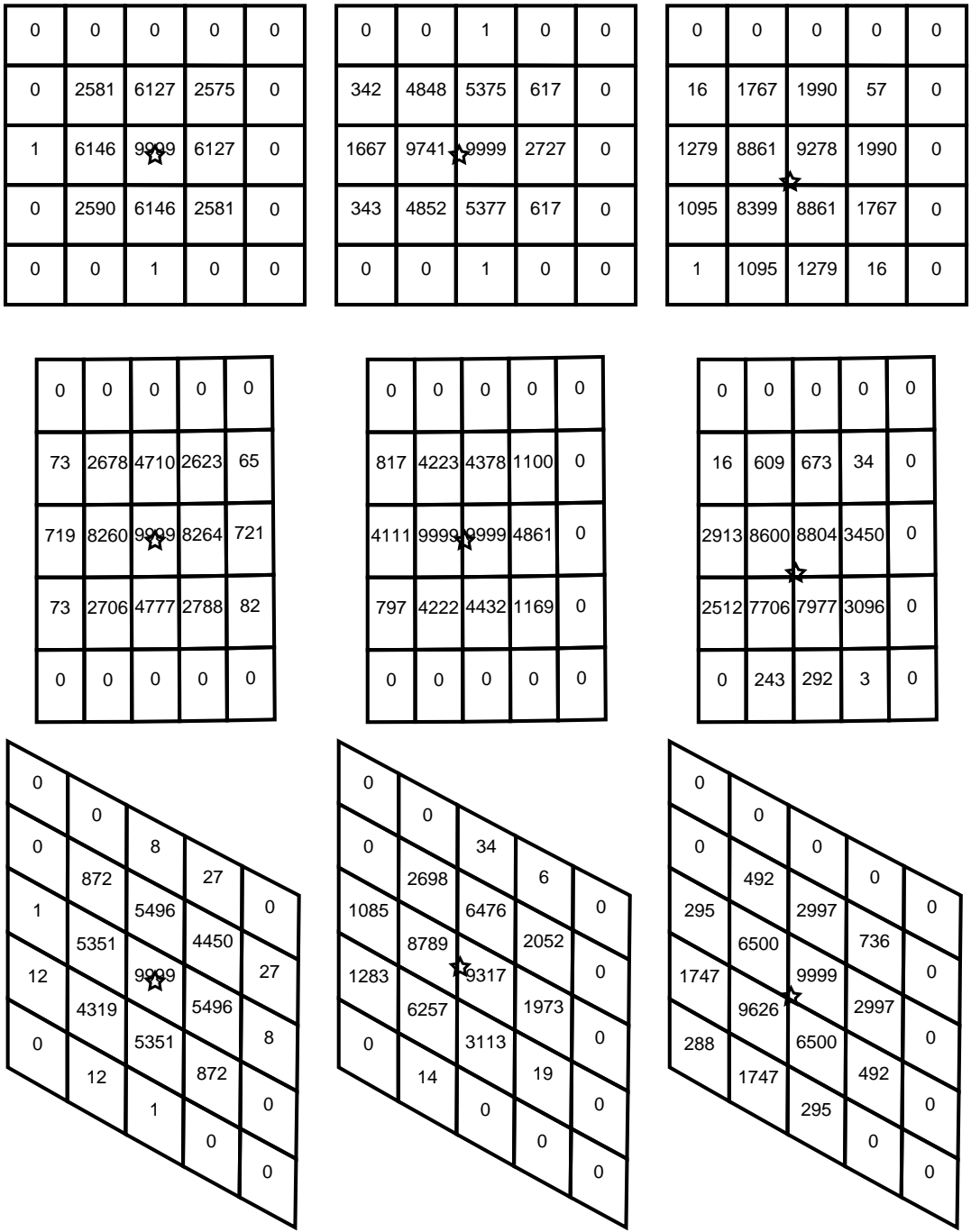}
\caption{%
The probability of a star at various locations within the DIRBE pixel grid
contributing to the signal in various pixels.  The top row is near the
center of a cube face, the middle row is near an edge of a cube face, and
the bottom row is near a corner of a cube face.  The numbers within each
pixel are $9999 p_{ij}$.
\label{fig:kmo}}
\end{figure}

\begin{figure}[tbp]
\plottwo{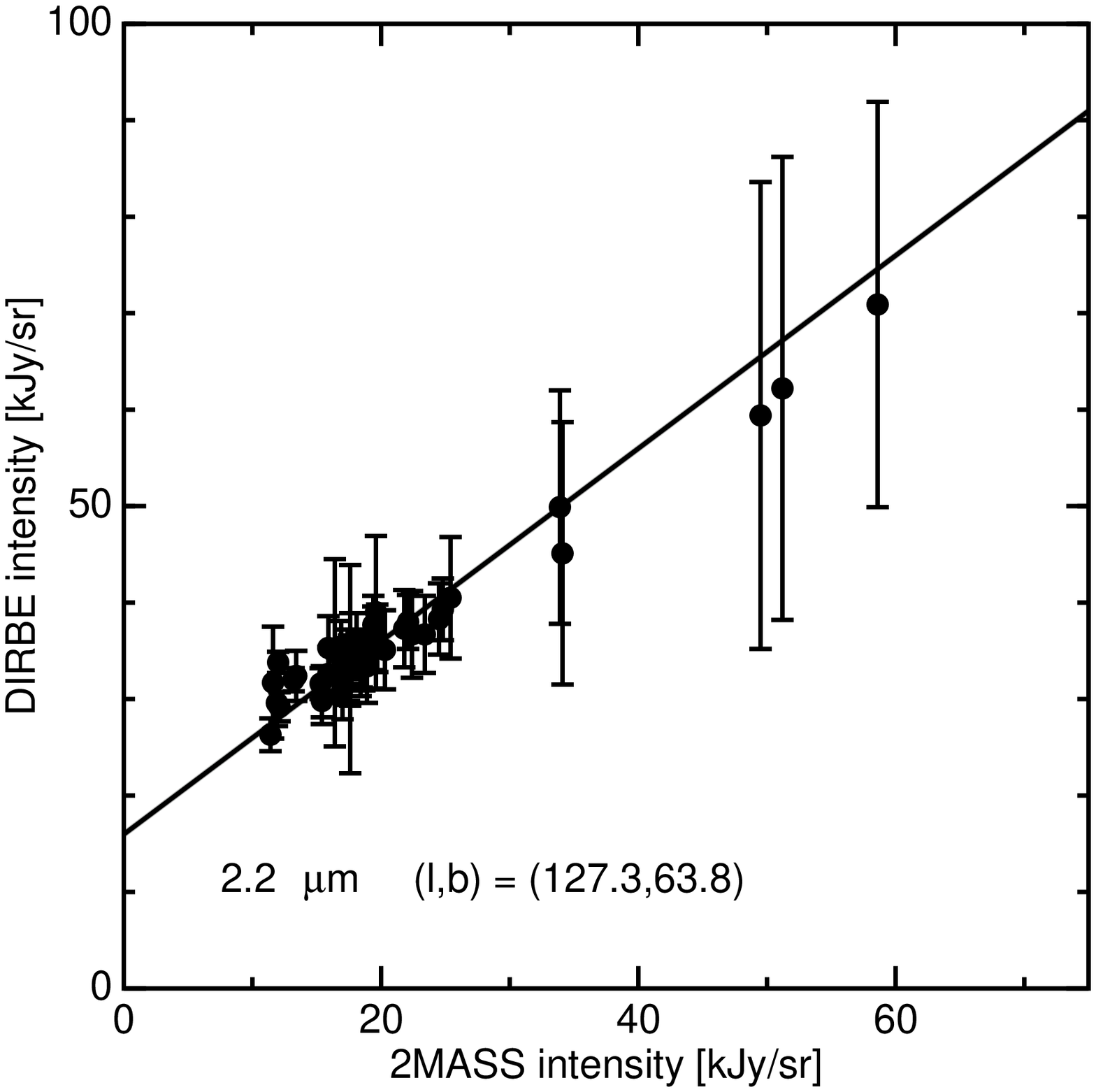}{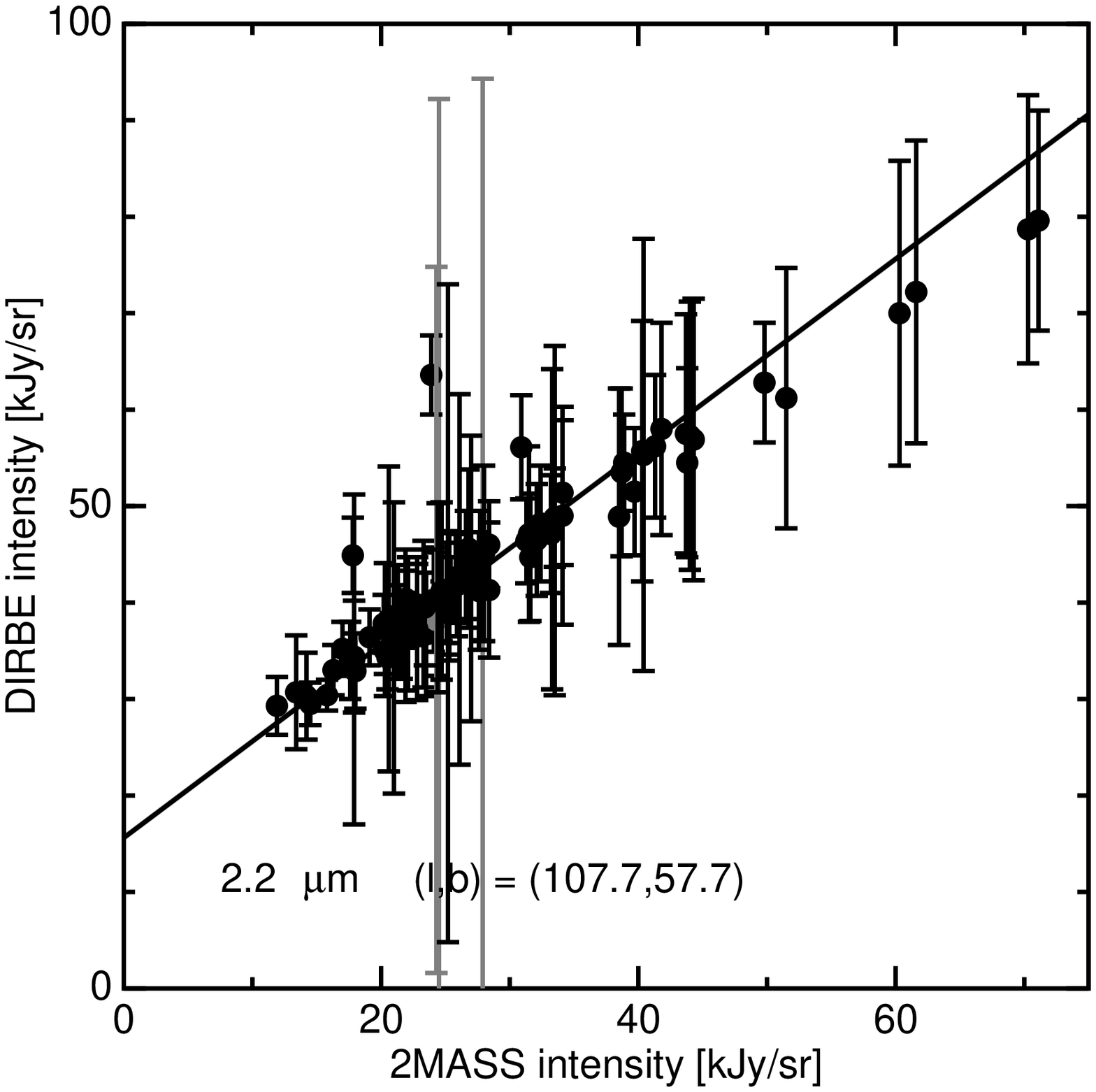}
\caption{%
Correlations of the DIRBE intensities \vs\ intensities 
computed from 2MASS fluxes in two Northern DIRBE dark spots.
\label{fig:KN}}
\end{figure}

\begin{figure}[tbp]
\plottwo{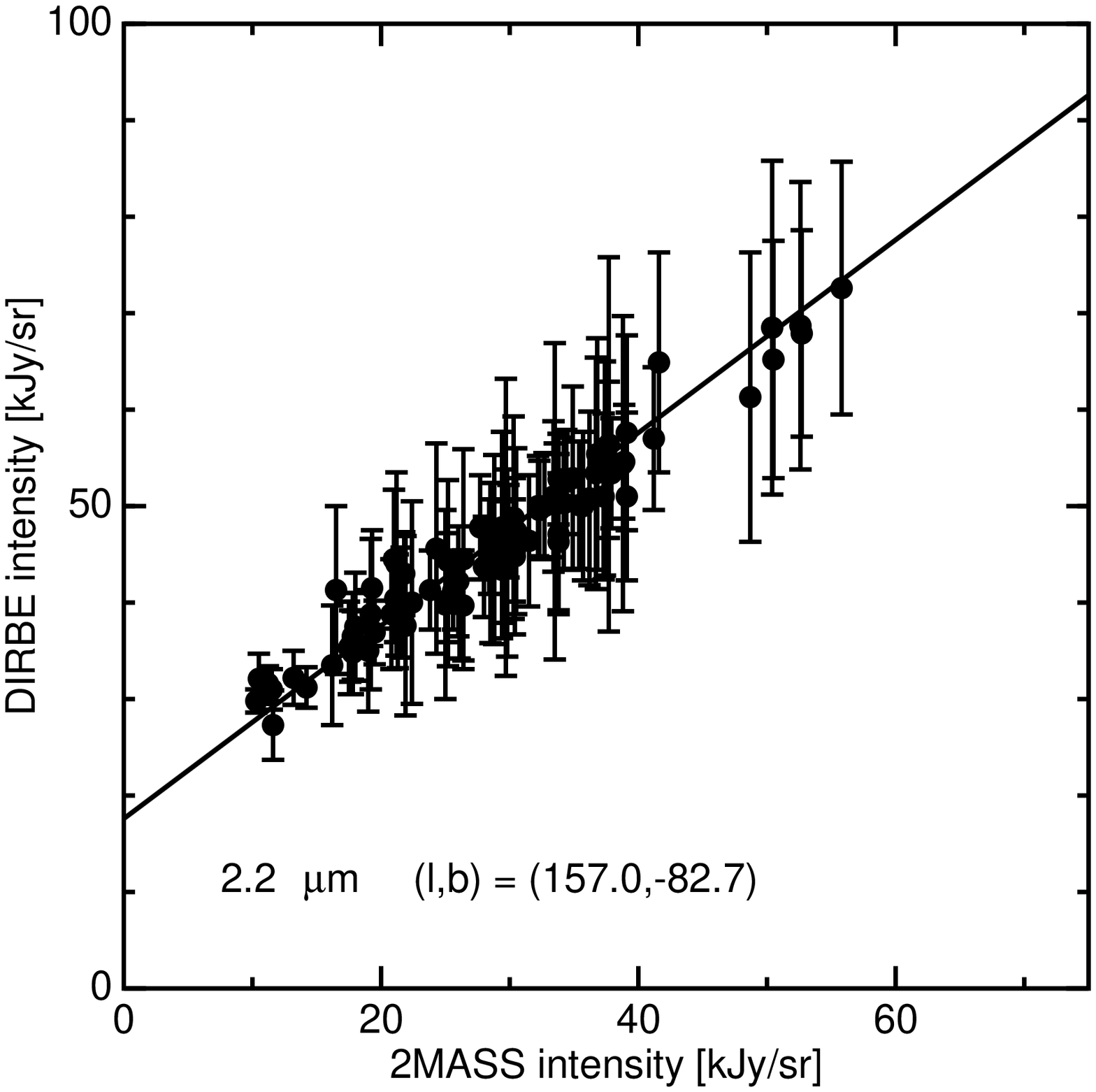}{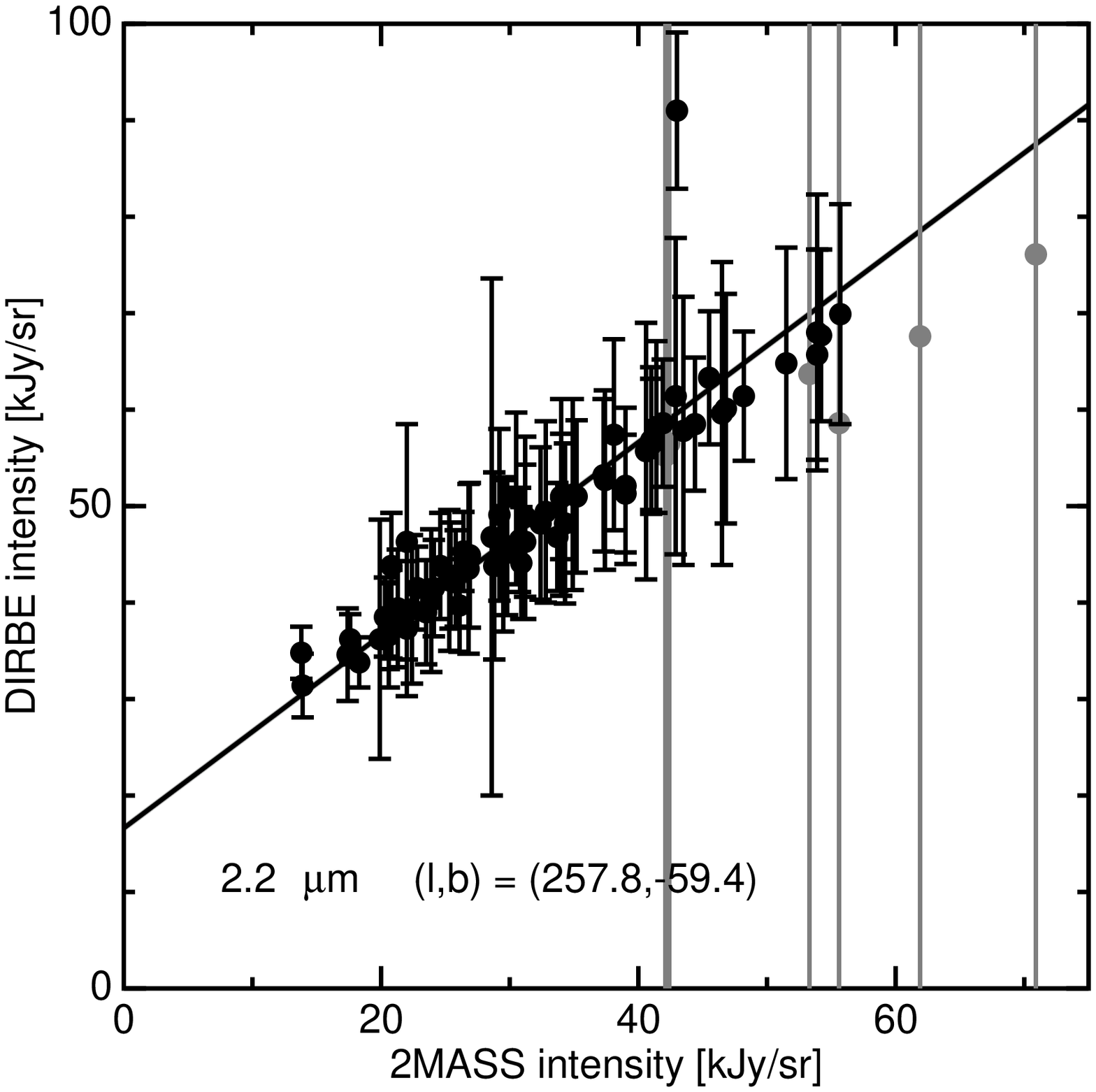}
\caption{%
Correlations of the DIRBE intensities \vs\ intensities 
computed from 2MASS fluxes in two Southern DIRBE dark spots.
\label{fig:KS}}
\end{figure}

The DIRBE data for the $i^{th}$ pixel is $D_i$, and should be the sum of
the zodiacal light, $Z_i$; the cataloged stars, $B_i$; the faint stars, $F_i$;
and the CIRB, $C$.  The cataloged star contribution is found using the method
of \citet{GWC00} on all stars brighter than $K = 14$.  
In this method, the DIRBE beam center is assumed to be uniformly distributed
within the area of the $i^{th}$ pixel, and the DIRBE beam orientation 
is assumed to be uniformly distributed in position angle.
Under these assumptions, the probability that the $j^{th}$ star contributes
to the signal in the $i^{th}$ pixel is $p_{ij}$, and the cataloged star
contribution is
\be
B_i = \Omega_b^{-1} \sum_j p_{ij} S_j
\label{eq:Bi}
\ee
where $S_j$ is the flux of the $j^{th}$ star and $\Omega_b$ is the solid angle
of the DIRBE beam.%
\footnote{\protect\cite{GWC00} used $\Omega_p$ instead of
$\Omega_b$ in Equation \protect\ref{eq:Bi}: this is appropriate for $p$'s
normalized to the total flux as in Table 5 of \citet{WR00}, but for $p$'s
normalized to a peak of unity as in Figure \protect\ref{fig:kmo} the beam solid
angle must be used for the normalization of Equation \protect\ref{eq:Bi}.}
Figure \ref{fig:kmo} shows the probability $p_{ij}$ for stars located in the
center, near an edge, or near a corner of a pixel located
near the center, an edge, or a corner of a cube face in the quadrilateralized
spherical cube pixel scheme.
The standard deviation of the bright star contribution is given by
\be
\sigma^2(B_i) = \Omega_b^{-2} \sum_j 
[p_{ij}(1-p_{ij}) + p_{ij}^2 (0.1+(0.4\ln10)^2\sigma^2(m_j))]S_j^2.
\label{eq:sigB}
\ee
The first term in the $[]$'s in Equation \ref{eq:sigB} is the ``flicker noise''
caused by a star that is on the edge of the DIRBE beam.  The second term
represents the flux error for the $j^{th}$ star, and it includes a generous
allowance for variability between the time of the DIRBE observations and the
the time of the 2MASS observations: the ``0.1'' corresponds to $\sigma = 0.34$
magnitudes.  This standard deviation is shown by the error bars in
Figures \ref{fig:KN} and \ref{fig:KS}.  The actual noise on the DIRBE data is
negligible: $1\;\kJysr$ at $1.25\;\um$ and $1.2\;\kJysr$ at $2.2\;\um$
\citep{HAKDO98}.
The uncertainty in the DIRBE zero level, or offset, is also negligible
\citep{HAKDO98}.
The CIRB should be 
isotropic, and $F$ should vary only slightly within a 2\deg\ radius of a 
dark spot.   But both the zodi-subtracted data, $\mbox{DZ}_i$, and the
bright star contribution fluctuate greatly from DIRBE beam to DIRBE
beam due to the confusion noise caused by overlapping stars.
Figures \ref{fig:KN} and \ref{fig:KS} 
show plots of $\mbox{DZ}_i$ \vs\ $B_i$
for each of the four fields.  
The gray points with large error bars are pixels with a bright star at the
edge of the beam.  These bright stars usually saturate the 2MASS survey,
and are assigned a nominal magnitude of 4 and a nominal flux error of a
factor of ten.  This large flux error guarantees that the pixels affected by
saturated stars have no effect on the subsequent analysis.
The lines have unit slope with intercepts
determined using a weighted median procedure.  The average values of
the data D$_i$, the zodi-subtracted data DZ$_i$, and the derived intercepts
DZ(0) for each of the four fields are given in Table \ref{tab:K}.

\begin{figure}[tbp]
\plotone{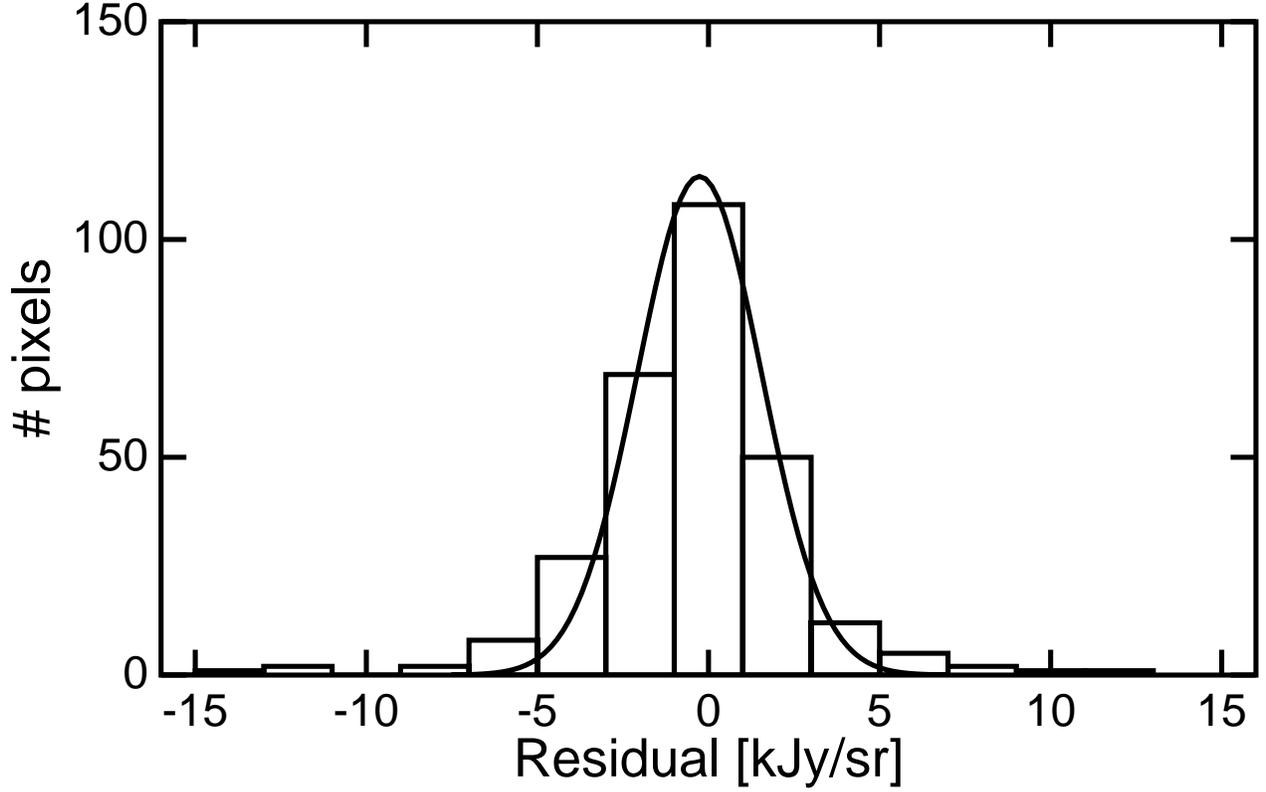}
\caption{%
Histogram of $\mbox{DZ}_i-B_i-\mbox{DZ(0)}$ for the four dark spots
combined at 2.2 \um.
The solid Gaussian
curve is fit to the highest three bins 
of the histogram.
\label{fig:Kresid}}
\end{figure}

The 2MASS $K_s$ magnitudes were converted to DIRBE fluxes using the
$F_\circ(K) = 614\;\mbox{Jy}$ derived by \citet{GWC00} from the median
of determinations using seven bright red stars.  This assumes that
there is no significant $K_s-K$ color difference between the red
calibration stars used by \citet{GWC00} and the stars in the dark spots.
The 2MASS observations saturate [even in the first read] on stars brighter
than $5^{th}$ magnitude.  At this level, the DIRBE data are still substantially
affected by confusion noise, so a direct comparison of DIRBE and 2MASS
on the same stars will not give a high precision result,
but the overall agreement between the unity slope lines 
and the data in Figures \ref{fig:KN} and \ref{fig:KS} shows
that a direct DIRBE to 2MASS comparison is consistent with
the \citet{GWC00} calibration.
Figure~\ref{fig:Kresid} shows a histogram of the values
$\mbox{DZ}_i-B_i-\mbox{DZ(0)}$ for the four dark spots combined.
Note that the standard deviations derived from the three
highest bins of these histograms of the individual histograms
are 1.27, 1.93, 2.82, and 2.93~\kJysr, while the standard deviation
of the Gaussian fit in Figure~\ref{fig:Kresid} is 1.81~\kJysr.
Since this histogram includes the DIRBE detector noise,
any small scale errors in the zodiacal light model,
any small scale structure in the faint star contribution,
a contribution from stellar variability,
and a calibration error contribution
in addition to any real extragalactic fluctuation, one can take
$1.81\;\kJysr = 2.47\;\nWmmsr$ as an upper limit on the extragalactic
fluctuation $\delta C_{rms}$.


\begin{table}[tbp]
\begin{center}
\begin{tabular}{rrrrrrrrr}
\hline
$l$ & $b$  & $\beta$ & $r$ & $N_{pix}$ &
	$\langle \mbox{D}_i\rangle $& $\langle \mbox{DZ}_i \rangle $  
	& DZ(0) & CIRB \\
$[^\circ]$ & $[^\circ]$  & $[^\circ]$ & $[^\circ]$ &  &
	 [kJy/sr] & [kJy/sr] &  [kJy/sr] & [kJy/sr] \\
\hline
127.3 &  63.8 &  50.8 & 1.5 & 44 & 139.65  & 36.77 & 15.98 & 14.50 \\
107.7 &  57.7 &  61.0 & 2.0 & 86 & 138.85  & 44.61 & 15.62 & 13.85 \\
157.0 & -82.7 & -26.0 & 2.0 & 86 & 214.58  & 46.22 & 17.60 & 16.20 \\
257.8 & -59.4 & -57.9 & 1.9 & 78 & 157.73  & 52.78 & 16.62 & 14.69 \\
\hline
\end{tabular}
\end{center}
\caption{Locations, sizes and intensities of the 4 DIRBE dark spots
at 2.2 $\mu$m.\label{tab:K}}
\end{table}

Stars fainter than $K = 14$ contribute a small amount which must be
subtracted from the intercepts.  This contribution was evaluated using the
\citet{WCVWS92} star count model.  But \citet{WR00} and \citet{GWC00}
find that this model overpredicts high latitude star counts 
by 10\% in the $6 < K < 12$ range.  
After applying this 10\% correction, which assumes that the 
same ratio applies to $K > 14$, the faint star corrections are
$F = 1.58$, 1.87,
1.50, and 2.03 \kJysr\ in the four fields.  
An uncertainty of 10\% of the total model prediction is assigned to this
correction, and listed in Table~\ref{tab:budget} under ``Faint Source''.
Galaxies brighter than $K =
14$ may be subtracted incorrectly, and their contribution should be added
back into the CIRB.  Galaxies with $K < 14$ add up 
to 0.35 \kJysr, according to the empirical fits of \citet{GCW93}.
A fraction of these galaxies will be in the 2MASS PSC and, since galaxies
should not be subtracted,
these incorrectly subtracted galaxies should be added back to the CIRB.  
The fluxes in the 2MASS Extended Source Catalog 
objects with $K_s < 14$ add up to 0.25 \kJysr, indicating 
that the correction for galaxies in the PSC should be on the order of
$0.1$~\kJysr.  In addition, 30\% of the
Extended Source Catalog objects are coincident with PSC objects suggesting
that the compact galactic nuclei -- which are in the PSC but should not be
subtracted from the CIRB -- account for 0.1 \kJysr.
Thus the CIRB estimates shown in Table \ref{tab:K} are given by
$\mbox{DZ(0)}-\mbox{F}+0.1\;\kJysr$.
An uncertainty of 100\% of this correction is included in 
Table~\ref{tab:budget} under ``Galaxies''.

Note that the statistical uncertainties in the intercepts are all $\leq
0.4\;\kJysr$ and thus negligible compared to systematic errors in the
interplanetary dust model.  \citet{GWC00} adopt an uncertainty of 5\% of
the zodiacal intensity at the ecliptic poles, and this gives an
uncertainty of $\pm 3.79\;\kJysr$.  
The effect of a $\pm 10\%$ calibration error between the DIRBE flux scale
and the 2MASS magnitude scale would be a systematic $\pm 2.58\;\kJysr$
change in the CIRB.
The precision of the \citet{GWC00} calibration of the DIRBE \Kband\ fluxes to 
the standard infrared magnitude is $\pm 2.1\%$.  
This DIRBE flux calibration agreed with \citet{AOWSH98}
calibration to better than $1\%$.  Thus a $10\%$ uncertainty in the DIRBE
{\it vs.} 2MASS calibration appears to be conservative.
This uncertainty is listed 
in Table~\ref{tab:budget} under ``Calibration''.
The mean of the CIRB estimates in
Table \ref{tab:K} is $14.79 \pm 0.51\;\kJysr$.  This
standard deviation of the mean of the 4 fields is listed in
Table~\ref{tab:budget} under ``Scatter''.
Finally, the largest uncertainty is the zodiacal light
modeling uncertainty.
Adding the errors in Table~\ref{tab:budget} in quadrature
gives a result of \KkJy\ or \KnW.

\begin{table}[tbp]
\begin{center}
\begin{tabular}{lrr}
\hline
Component & $1.25\;\um$ & $2.20\;\um$ \\
\hline
Scatter         &          1.49 &             0.51 \\
Faint Source    &          0.34 &             0.18 \\
Galaxies        &          0.05 &             0.10 \\
Calibration     &          3.10 &             2.58 \\
Zodiacal        &          5.87 &             3.79 \\
Quadrature Sum  &          6.81 &             4.62 \\
\hline
\end{tabular}
\end{center}
\caption{Error budget for the CIRB.\label{tab:budget}}
\end{table}

\section{J Band}

\begin{figure}[tbp]
\plotone{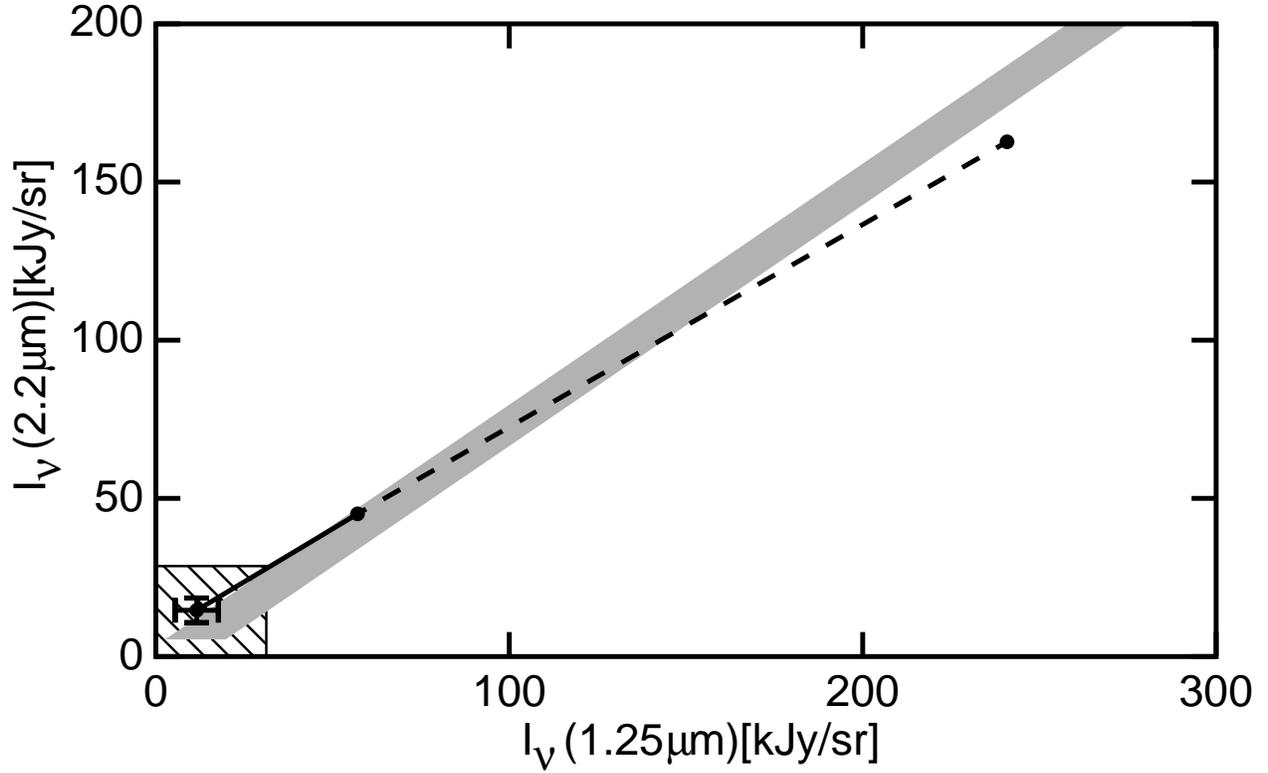}
\caption{%
$J$ \vs\ $K$ intensities, averaged over the four dark spots, 
with the highest point showing the raw data, a dashed line showing the
zodiacal model subtraction, and a solid line showing the star 
subtraction yielding the CIRB -- the point with errorbars.  
The \protect\citet{HAKDO98} upper limits are shown 
as the cross-hatched region, while the gray band shows 
the \protect\citet{DA98} correlation.\label{fig:CBR-JK}}
\end{figure}

\begin{table}[tbp]
\begin{center}
\begin{tabular}{rrrrrrr}
\hline
$l$ & $b$  &
	$\langle \mbox{D}_i\rangle $& $\langle \mbox{DZ}_i \rangle $  
	& DZ(0) & F & CIRB \\
$[^\circ]$ & $[^\circ]$  &
	 [kJy/sr] & [kJy/sr] &  [kJy/sr] &  [kJy/sr] & [kJy/sr] \\
\hline
127.3 &  63.8 & 211.19 & 48.23 & 16.67 & 3.08 & 13.64 \\
107.7 &  57.7 & 205.92 & 58.86 & 16.14 & 3.62 & 12.57 \\
157.0 & -82.7 & 314.80 & 51.48 & 10.59 & 2.94 &  7.70 \\
257.8 & -59.4 & 231.47 & 70.01 & 18.13 & 3.92 & 14.26 \\
\hline
\end{tabular}
\end{center}
\caption{Intensities of the 4 DIRBE dark spots
at 1.25 $\mu$m.\label{tab:J}}
\end{table}

For the \Jband\ the contribution from stars with $J < 14$
{\em and} $K < 14$ was calculated on a pixel by pixel basis.
This dual wavelength magnitude selection is essentially equivalent
to a simple $J < 14$ selection.  There are very few stars in high Galactic
latitude fields with color $J\>-\>K < 0$.  Tests in three fields using deeper
samples from the 2MASS catalog show that imposing the $K < 14$
cut on a $J < 14$ sample reduces the intensity by $< 0.3$\%.
Table \ref{tab:J} gives the photometric quantities for the four dark fields
at 1.25 \um.
The $(10 \pm 10)\%$ correction of the faint star contribution derived 
from \Kband\ star counts has been applied to $F$.
Assuming a color of $J-K = 1$ for galaxies gives an estimate of
$0.05\;\kJysr$ for the improperly subtracted faint galaxy contribution.
This has been added to the CIRB estimates in the table.
The mean of the CIRB estimates is $12.04 \pm 1.49\;\kJysr$,
and this would change by $\mp 3.10\;\kJysr$ for $\pm 10\%$ changes in 
the flux of 1512~Jy for $0^{th}$ magnitude at $1.25\;\um$ used in this paper.
This value is the median of calibrations based on $\beta$ And, $\alpha$ Tau,
$\alpha$ Aur, $\alpha$ Boo, $\alpha$ Her, and $\beta$ Peg.
This calibration has an uncertainty of $\pm 2\%$ and differs from the
\citet{AOWSH98} calibration by -2.4\%.
A $\pm 10\%$ uncertainty in the calibration of DIRBE {\it vs.} 2MASS has
been adopted giving the error
$\pm 3.10\;\kJysr$ listed in Table~\ref{tab:budget} .
The systematic error due to interplanetary dust modeling 
is 5\% of the ecliptic pole zodiacal intensity or $\pm 5.87\;\kJysr$.  
Adding the errors in quadrature gives a result of \JkJy\ or
\JnW.  This is obviously
not a significant detection due to the large uncertainty in the zodiacal
foreground.  However, a $2\sigma$ upper limit is $\nu I_\nu < 62\;\nWmmsr$
which is a slight improvement on \citet{HAKDO98}.

The width of the combined histogram of the residuals
for the \Jband\ data is $2.32\;\kJysr$ or $5.6\;\nWmmsr$
which gives an upper limit on extragalactic fluctuations since it also
includes detector noise and star subtraction errors.


A CIRB of \JkJy\ at 1.25 \um\ is fainter than the prediction of \citet{DA98},
whose correlation gives $23.5 \pm 8.6\;\kJysr$ for this paper's 
\Kband\ CIRB. Given the combined uncertainties in the difference, this is 
$< 1.2 \sigma$ higher than the result in this paper.

\section{Discussion}

\begin{figure}[tbp]
\plotone{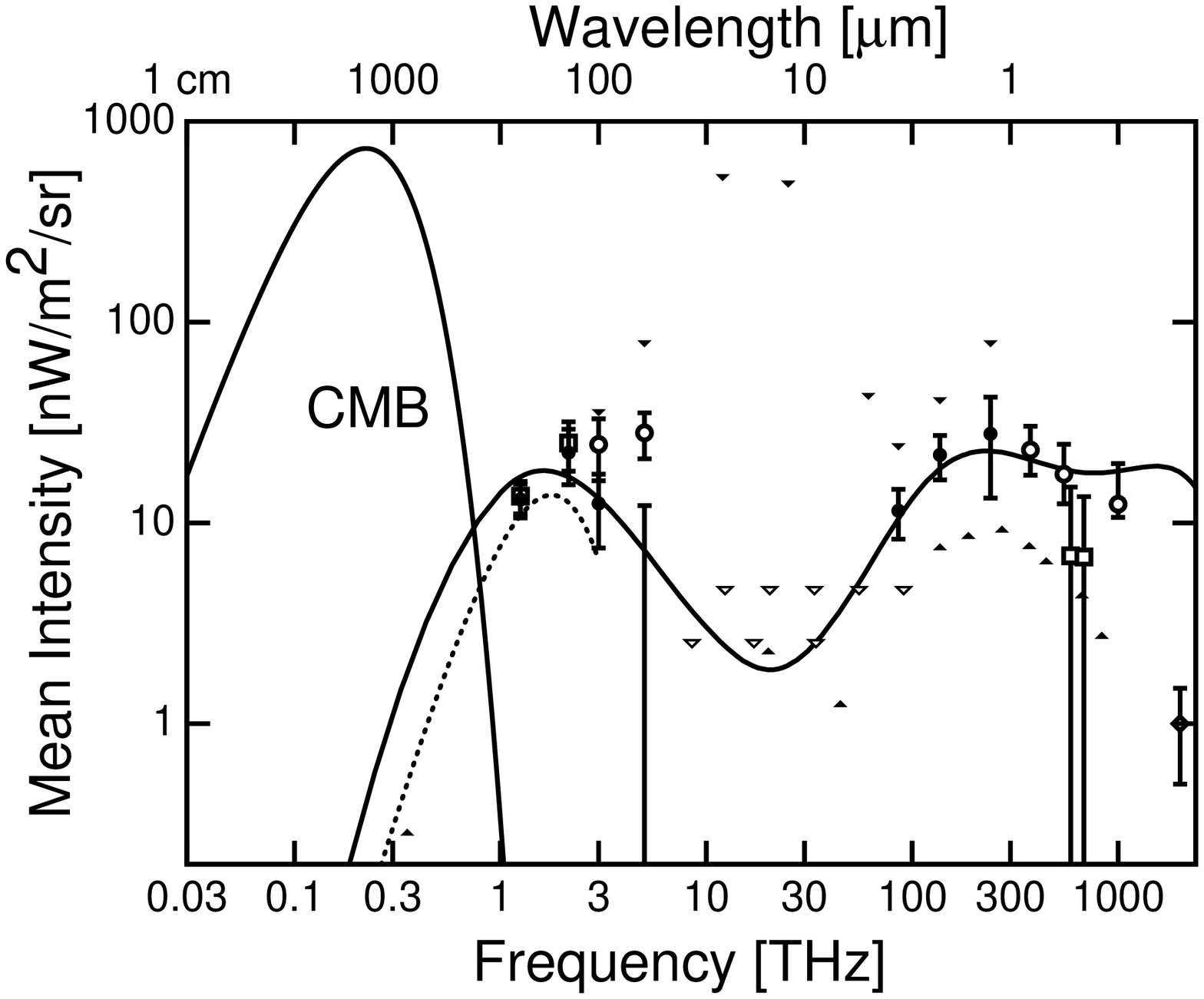}
\caption{%
Comparison of CIRB values to previous determinations and upper limits.
Lower limits from source counts from \protect\citet{SIBK99} at 850 \um,
\protect\citet{1997fisu.conf..159F} at 15 \& 6.7 \um, and 
\protect\citet{MP00} at 2.2 to 0.3 \um.
Solid upper limits from \protect\citet{HAKDO98},
open upper limit symbols using
$\gamma$-rays from \citet{FMMRW98} and \citet{SF98}.
Open squares at 240 \& 140 \um\ from \protect\citet{HAKDO98}, open circles
at 100 \& 60 \um\ from \citet{FDS00}, while the filled circle
far IR data points are the
\citet{HAKDO98} results modified by using this paper's zodiacal model.
Dashed curve is from \protect\citet{FDMBS98}.
Filled circles from 3.5 to 1.25 \um\ are an average of 
\protect\citet{GWC00}, \protect\citet{WR00} and this paper.
Open circles from 0.8 to 0.3 \um\ are from \protect\citet{1999hrug.conf..487B}.
Open squares are from \protect\citet{DWW77} and \protect\citet{T83}.
The open diamond at 0.15 \um\ is from \protect\citet{HBM91}.
\label{fig:COIBR-prev}}
\end{figure}

Subtracting the 2MASS catalog from the zodi-subtracted
DIRBE data yields a statistically
significant, isotropic background at 2.2 \um\ of \KkJy\ which is consistent 
with the earlier results from \citet{GWC00} $(16.4 \pm 4.4\;\kJysr)$ 
and \citet{WR00} $(16.9 \pm 4.4\;\kJysr)$ within the
systematic error associated with the modeling the zodiacal dust cloud.
Averaging the results of \citet{GWC00}, \citet{WR00} and this paper gives
a CIRB at 2.2 \um\ of $16 \pm 4\;\kJysr$.  This averaging has not 
reduced the estimated error which is dominated by systematic effects
that affect all three results equally.
The foreground
due to interplanetary dust at 1.25 \um\ is too large to allow a CIRB
detection, but an improved upper limit is found.  
Note that the Zodi-Subtracted Mission Average maps which used the
\citet{KWFRA98} zodiacal light model give a CIRB that is 
13.75 \kJysr\ larger at 1.25 \um\ and 6.08 \kJysr\ larger at 2.2 \um\ than
results obtained here using
the zodiacal light model described in \citet{Wr98} and \citet{GWC00} based on
the very strong no-zodi principle
of \citet{Wr97}.
Figure \ref{fig:CBR-JK}
shows a plot of the \Jband\ intensity \vs\ \Kband\ intensity averaged
over the four
dark spots analyzed in this paper.  Three values are plotted:
the average total intensity $\langle \mbox{D} \rangle$, the average
zodi-subtracted intensity $\langle \mbox{DZ} \rangle$, and the CIRB estimates.
The \citet{HAKDO98} upper limits on the CIRB,
the \cite{DA98} correlation and the $1 \sigma$ error 
bars from this paper are shown as well.  This figure emphasizes the
large subtractions that are involved in determining the CIRB from data
taken 1~AU from the Sun: the zodiacal light is about 16 times larger than the
CIRB at 1.25 \um\ and 8 times larger than the CIRB at 2.2 \um.
Galactic stars are a problem in the large DIRBE beam, but in the selected
dark spots the effect of stars is 4 times less than that of the zodiacal
light.

\citet{1999hrug.conf..487B} has measured the optical extragalactic
background light and obtained results at $\lambda = 0.8,\;0.55,\;
\mbox{\&}\;0.3\;\um$ which are consistent with a reasonable
extrapolation through the uncertain \Jband\ result found here,
as shown in Figure \ref{fig:COIBR-prev}.
Both \citet{1999hrug.conf..487B} and this work face challenging and uncertain
corrections for the zodiacal light, but the two papers use very different
techniques and should not have systematic errors in common.  Thus the
lack of a discontinuity in the spectrum between 0.8 and 1.25 \um\
is an indication in favor of the background level reported here.
The model shown in Figure \ref{fig:COIBR-prev} is the 
$\Lambda$CDM-Salpeter model from
\citet{PBSN99} which appears to fit the observed far IR to near IR
to optical ratios.
But the model was multiplied by 1.84 to match the level
of the observed background.

\acknowledgments

The {\sl COBE} datasets were developed by the NASA Goddard Space Flight
Center under the guidance of the COBE Science Working Group and were
provided by the NSSDC.  This publication makes use of data products from the
Two Micron All Sky Survey, which is a joint project of the University of
Massachusetts and the Infrared Processing and Analysis Center, funded by the
National Aeronautics and Space Administration and the National Science
Foundation.



\end{document}